\documentstyle[12pt]{article}
\begin{document}
%July 14, 1995
%\baselineskip=24pt
\newcommand{\Lam}{\Lambda_{\scriptscriptstyle {\rm \overline{MS}} }}
\newcommand{\Lamm}[1]{\Lambda {\vspace{-2pt} \scriptscriptstyle ^{^{\rm #1}}
    _{\rm \overline{MS}} }}
\newcommand{\dsp}{\displaystyle}
\newcommand{\dfr}[2]{ \displaystyle\frac{#1}{#2} }
\newcommand{\Lag}{\Lambda \scriptscriptstyle _{ \rm GR} }
\newcommand{\pa}{p\parallel}
\newcommand{\pe}{p\perp}
\newcommand{\paa}{p'\parallel}
\newcommand{\pee}{p'\perp}
\renewcommand{\to}{\rightarrow}
\renewcommand{\baselinestretch}{1.5}
\vspace{-20ex}
\begin{flushright}
\vspace{-3.0ex}
    \it{AS-ITP-95-23} \\
\vspace{-2.0mm}
       \it{July, 1995}\\
%\vspace{-2.0mm}
\vspace{5.0ex}
\end{flushright}
\centerline{\Large\bf Implications on two Higgs doublet models}
\vspace{2mm}
\centerline{\Large\bf from the latest $b\to s\gamma$ measurement and
top discovery}
\vspace{6.4ex}
\centerline{\large\bf  Chao-Hsi Chang\footnote{
E-mail: zhangzx@itp.ac.cn.} and
Caidian L\"u\footnote{ E-mail: lucd@itp.ac.cn.}}
\vspace{4mm}
\vspace{3.5ex}
\centerline{\sf CCAST (World Laboratory), P.O.Box 8730, Beijing 100080,
China,}
\centerline{\sf and}
\centerline{\sf Institute of Theoretical Physics, Academia Sinica,
P.O.Box 2735, Beijing 100080, China.\footnote{Mailing address.}}

\vspace{3ex}
\begin{center}
\begin{minipage}{5in}
\centerline{\large\bf 	Abstract}
\vspace{1.5ex}
\small
{The constraints on the two Higgs doublet model
from the new experimental bounds of $b\to s\gamma$ by CLEO and the
latest published value of the top quark mass by CDF and D0
are reanalyzed
with the effective Lagrangian covering the full QCD corrections
from the energy scale of top quark to that of bottom.
The reanalysis result shows that the constraints become more
stringent than that of the earlier analysis,
i.e. a bigger region of the parameter space of the model is ruled out.}

\vspace{4mm}
\end{minipage}
\end{center}

{\bf PACS} number(s):14.80.Cp, 12.60.Fr, 12.38.Bx

\newpage

It is known that the experimental bounds of $b\to
s\gamma$ set very strong constraints on the two Higgs doublet
model (2HDM), a minimal extension of the Standard Model (SM).
In additional to searching for the neutral Higgs of minimal SM,
phenomenologically to investigate
possible extensions of SM is also another hot topic in particle
physics, thus to apply the latest experimental results of the measurement
on $b\to s\gamma$\cite{cleo2} and the newly discovery of top quark\cite{t1}
to reexamine the constraints on the 2HDM so as to upgrade
the allowed values of the model parameters is no doubt always
to be interesting.

Reviewing the earlier analysis\cite{s1}, one would find that
the QCD correction effects owing to the change of the energy scale
from top quark's down to that of $W$ boson were ignored.
Indeed this piece of QCD correction itself is not great, but we
treat them seriously, and finally find it being not negnigible
since this correction affects the constraints on the two
Higgs doublet model sizable in the report.

Let us first recall the necessary formulas here for later use
(mainly from ref.\cite{lcd1}).
There are two models for 2HDM
to avoid tree-level flavor changing neutral currents (FCNCs).
The first (Model I) is to allow only one of the two Higgs doublets to
couple to both types, u-type and d-type, of quarks\cite{Hab} but
the other doublet is totally forbidden by certain discrete symmetry.
The second (Model II) is to arrange as that
one Higgs doublet couples to u-type quarks while the other
couples to d-type quarks respectively due to a different
discrete symmetry\cite{Gla}.
It is of interest to note that the Model II,
as a natural feature, occurs in such a theory as that with
supersymmetry or with a Peccei-Quinn type of symmetry.

The piece of the 2HDM Lagrangian for the charged Higgs to the quarks is
\begin{eqnarray}
 {\cal L}&= &
    \frac{1}{\sqrt{2}} \frac{\mu^{\epsilon/2}g_2}{M_W}
      \left[\left(\frac{v_2}{v_1}\right)
	\left(\begin{array}{ccc} \overline{u} & \overline{c} &
	      \overline{t} \end{array}\right)_R M_U V
	      \left(\begin{array}{c} d \\ s \\ b \end{array}\right)_L
	     -\xi
		\left(\begin{array}{ccc} \overline{u} & \overline{c} &
	     \overline{t} \end{array}\right)_L V M_D
	    \left(\begin{array}{c} d \\ s \\ b \end{array}\right)_R
	     \right] H^+ \nonumber\\
	&  +&h.c. \nonumber\\
\end{eqnarray}
where V represents the $3 \times 3$ unitary Cabibbo-Kobayashi-Maskawa
(CKM) matrix,
$M_U$ and $M_D$ denote the diagonalized quark mass matrices, the subscript
$L$ and $R$ denote left-handed and right-handed quarks, respectively.
For Model I, $\xi={v_2}/{v_1}$; while for Model II, $\xi=-{v_1}/{v_2}$.
And $v_1$, $v_2$ are the magnitude of the vacuum expectation values of
two Higgs doublets, respectively.

The relevant effective Hamiltonian after integrating out the heavy top
freedom is:
\begin{equation}
{\cal H}_{eff}=2 \sqrt{2} G_F V_{tb}V_{ts}^*\displaystyle \sum _i
C_i(\mu)O_i(\mu). \label{eff}
\end{equation}
The coefficients $C_i(m_t )$ can be calculated from
matching conditions at $\mu=m_t$, and $C_i(\mu )$ can be obtained from
their renormalization group equation(RGE):
\begin{equation}
\mu \frac{d}{d\mu} C_i(\mu)=\displaystyle\sum_{j}(\gamma^{\tau})_{
ij}C_j(\mu),\label{ren}
\end{equation}
where $(\gamma^{\tau})_{ij}$ is the anomalous dimension matrix of the
operators $O_i$. If integrating out the $W$ boson freedom
further, once more
six relevant four-quark operators will be added\cite{lcd1}.
We do not repeat all the operators   here but only write
down the most relevant operators:
\begin{eqnarray}
O_7&=&(e/16\pi^2) m_b \overline{s}_L \sigma^{\mu\nu}
	    b_{R} F_{\mu\nu},\nonumber\\
O_8&=&(g/16\pi^2) m_b \overline{s}_{L} \sigma^{\mu\nu}
	    T^a b_{R} G_{\mu\nu}^a.
\end{eqnarray}

The  effective Hamiltonian appears just below the W-scale as
\begin{eqnarray}
{\cal H}_{eff} && =\frac{4G_F}{\sqrt{2}} V_{tb}V_{ts}^*
	\displaystyle \sum_{i}
                C_i(M_{W}) O_i(M_{W})\nonumber\\
	& &\stackrel{EOM}{\rightarrow}
		\frac{4G_F}{\sqrt{2}} V_{tb}V_{ts}^*\left\{
		\displaystyle{\sum_{i=1}^{6} }C_i (M_{W})O_i + C_7(M_{W})
                O_7 +C_8(M_{W}) O_8 \right\}.
\end{eqnarray}

For completeness, the explicit expressions of
the coefficient of operators at $\mu=M_{W}$ are given\cite{lcd1},
$$ C_i(M_{W})=0,~~~for ~i=1,3,4,5,6;~~~ C_2(M_{W})=1,$$
\nopagebreak[1]
\begin{eqnarray}
C_{O_8}(M_{W}) &= & \left( \frac{\alpha _s (m_t)} {\alpha _s (M_W)}
	\right) ^{ \frac{14}{23} } \left\{ \frac{1}{2}C_{O_{LR}^1}(m_t)
-C_{O_{LR}^2}(m_t) +\frac{1}{2}C_{P_L^{1,1}}(m_t) \right.\nonumber\\
&&	\;\;\;\;\;\;\;\;\;\;\;\;\;\;\;\;\;\;
	\left.+\frac{1}{4}C_{P_L^{1,2}}(m_t)
	-\frac{1}{4}C_{P_L^{1,4}}(m_t)\right\}
	-\frac{1}{3} ,\label{c2}
\end{eqnarray}
\begin{eqnarray}
&\displaystyle{
C_{O_7}(M_{W}) = \frac{1}{3}\left( \frac{\alpha _s (m_t)} {\alpha _s (M_W)}
	\right) ^{ \frac{16}{23} } \left\{ C_{O_{LR}^3}(m_t)
	+8 C_{O_{LR}^2}(m_t) \left[1-\left( \frac{\alpha _s (M_W)}
{\alpha _s (m_t)} \right) ^{ \frac{2}{23} } \right]\right.}&\nonumber\\
&\displaystyle{	+\left[-\frac{9}{2} C_{O_{LR}^1}(m_t)
	-\frac{9}{2}C_{P_L^{1,1}}(m_t)
-\frac{9}{4}C_{P_L^{1,2}}(m_t) +\frac{9}{4}C_{P_L^{1,4}}(m_t)\right]
\left[1- \frac{8}{9} \left( \frac{\alpha _s (M_W)}
{\alpha _s (m_t)} \right) ^{ \frac{2}{23} } \right] }&\nonumber\\
&\displaystyle{	\;\;\;\left.-\frac{1}{4}C_{P_L^4}(m_t)
+\frac{9}{23} 16\pi^2 C_{W_L^1}(m_t) \left[1- \frac{\alpha _s (m_t)}
	{\alpha _s (M_W)} \right] \right \}
	-\frac{23}{36} }, &  \label{c3}
\end{eqnarray}
and together with the coefficients of operators at $\mu=m_t$,
$$C_{W_L^1} = \delta /g_3^2,$$
\begin{eqnarray}
C_{O_{LR}^1}&=&-\left(\frac{1+\delta}
	{2(1-\delta)^2}+\frac{\delta}{(1-\delta)^3}\log\delta\right)
	-\xi'
	\left( \frac{ 1 +x }{2(1-x)^2}
	+\frac{x}{(1-x)^3}\log x \right), \nonumber\\
C_{O_{LR}^2}&=&-\frac{1}{2} \left(\frac{1}{(1-\delta)}+
	\frac{\delta}{(1-\delta)^2}\log\delta\right)
  	-\xi'
	\left( \frac{ 1 }{2(1-x)}
	+\frac{x}{2(1-x)^2}\log x \right), \nonumber\\
C_{O_{LR}^3}&=& \left(\frac{1}{(1-\delta)}+
	\frac{\delta}{(1-\delta)^2}\log\delta\right)
	+\xi
	\left( \frac{ 1 }{1-x} +\frac{x}{(1-x)^2}\log x \right), \nonumber\\
C_{P_L^{1,1}}&=& C_{P_L^{1,3}}\;\;=\;\;
	\left(\frac{\frac{11}{18}+\frac{5}{6}\delta
	-\frac{2}{3}\delta^2 +\frac{2}{9} \delta ^3}{(1-\delta)^3}+
\frac{\delta+\delta^2-\frac{5}{3}\delta^3 +\frac{2}{3} \delta ^4}
	{(1-\delta)^4}\log\delta\right)\nonumber\\
  &&~~~~~~~~+\left(\frac{v_2}{v_1}\right)^2 \left(
	\frac{\frac{11}{18} -\frac{7}{18}x +\frac{1}{9}x^2 }{ (1-x)^3}
	+\frac{x-x^2+\frac{1}{3}x^3 }{ (1-x)^4 }\log x \right),\nonumber\\
C_{P_L^{1,2}}&=&\left(\frac{-\frac{8}{9}-\frac{1}{6}\delta
	+\frac{17}{6}\delta^2 -\frac{7}{9} \delta ^3}{(1-\delta)^3}+
	\frac{-\delta+\frac{10}{3}\delta^3 -\frac{4}{3} \delta^4}
	{(1-\delta)^4}\log \delta \right)\nonumber\\
  &&+\left(\frac{v_2}{v_1}\right)^2 \left(
	\frac{-\frac{8}{9} +\frac{29}{18}x -\frac{7}{18}x^2 }{ (1-x)^3}
	+\frac{-x+2x^2-\frac{2}{3}x^3 }{ (1-x)^4 }\log x \right),\nonumber\\
C_{P_L^{1,4}}&=&\left(\frac{\frac{1}{2}-\delta
	-\frac{1}{2}\delta^2 +\delta^3}{(1-\delta)^3}+
\frac{\delta-3\delta^2+2\delta^3}{(1-\delta)^4}\log\delta\right)\nonumber\\
  &&+\left(\frac{v_2}{v_1}\right)^2 \left(
	\frac{1 -x^2 }{ 2(1-x)^3}
	+\frac{x-x^2 }{ (1-x)^4 }\log x \right),\nonumber\\
C_{P_L^4}&=&\frac{1}{Q_b}\left(\frac{-\frac{1}{2}-5\delta
	+\frac{17}{2}\delta^2 -3\delta^3 }{(1-\delta)^3}+
	\frac{-5\delta +7\delta^2 -2\delta^3}{(1-\delta)^4}\log\delta
	+4\delta \log\delta \right)\nonumber\\
  &&-\frac{1}{Q_b}\left(\frac{v_2}{v_1}\right)^2 \left(
	\frac{1 -x^2 }{ 2(1-x)^3}
	+\frac{x-x^2 }{ (1-x)^4 }\log x \right).\label{coe}
\end{eqnarray}
Where $\delta = M_W^2/m_t^2$, $x=m_{\phi}^2 /m_t^2$;
With
$$\xi'=v_2^2/v_1^2,~~~~~ Model~ I, $$
$$\xi'=-1,~~~~~~~  Model~ II.$$
When $\xi =\xi'=0$, the above result (\ref{coe}) reduces to that of SM
 case\cite{lcd2}.

The running of the coefficients of operators from $\mu=M_W$ to $\mu=m_b$
was well described in refs.\cite{Grin,Gri,Cel,Mis,Ade,Ciu}. With the
running due to QCD, the coefficients of the operators at $\mu=m_b$ scale:
\begin{equation}
C_7^{eff}(m_b) = \eta^{16/23}C_7(M_W) +\frac{8}{3}
( \eta^{14/23}-\eta^{16/23} ) C_8(M_W)
+C_2(M_W) \displaystyle \sum _{i=1}^{8} h_i \eta^{a_i}.
\end{equation}
Here $\eta = \alpha_s(M_W) /\alpha_s (m_b)$,
$$ h_i =\left( \frac{626126}{272277}, -\frac{56281}{51730},
-\frac{3}{7}, -\frac{1}{14}, -0.6494, -0.0380, -0.0186, -0.0057 \right),$$
$$a_i = \left( \frac{14}{23}, \frac{16}{23}, \frac{6}{23}, -\frac{12}{23},
0.4086, -0.4230, -0.8994, 0.1456 \right).$$

Following refs.\cite{Grin,Gri,Mis,Ade,Ciu},
\begin{eqnarray}
BR(B \rightarrow X_s \gamma) /BR(B\rightarrow X_c e\overline{\nu})
\simeq\Gamma(b\rightarrow s\gamma)/\Gamma
(b\rightarrow ce\overline{\nu}).
\end{eqnarray}
Then
\begin{eqnarray}
\frac{BR(B \rightarrow X_s \gamma)}{BR(B
\rightarrow X_c e \overline{\nu})} \simeq \frac{|V_{ts}^*V_{tb}|^2}{
|V_{cb}|^2} \frac{6 \alpha_{QED}}{\pi g (m_c/m_b)}
|C_7^{eff}(m_b)|^2,
\end{eqnarray}
where the phase space factor $g(z)$ is given by:
\begin{equation}
g(z)=1-8z^2+8z^6-z^8-24z^4\log z,
\end{equation}
here we use $m_c/m_b=0.316$.
Afterwards one obtains the $B \rightarrow X_s
\gamma$ decay rate normalized to the quite well established
semileptonic decay rate $Br(B \to X_c e\overline{\nu} )$.
If we take experimental result $BR(B \to
X_c e\overline{\nu} ) =10.8\% $\cite{data}, the branching ratios of
$B \to X_s \gamma$ is found to be:
\begin{eqnarray}
BR(B \rightarrow X_s \gamma)
\simeq 10.8\%\times \frac{|V_{ts}^*V_{tb}|^2}{ |V_{cb}|^2}
\frac{6 \alpha_{QED}\;|C_7^{eff}(m_b)|^2}
{\pi g (m_c/m_b)}.
\end{eqnarray}

Note that in the above equations the top mass $m_t$ is kept as a parameter
precisely thus we may apply them to computing all the values with
various experimental $m_t$ as one needs.

To emphasize the consequences have been ignored in literature,
first of all we plot the coefficients of the most relevant
operators $O_7$ and $O_8$ at $\mu= M_W$ versus $\tan \beta$
with and without the QCD corrections of
the energy scale running from $m_t$ (with the latest experimental value)
to $M_W$ in Fig.1 for Model I and Model II respectively.
Here $m_b=4.8$GeV, $M_W=80.22$GeV and the QCD coupling $\alpha_s(m_Z)=
0.125$ are taken\cite{data}.
Owing to the mixing of all the relevant operators being small, one may
see the fact that the effects of the QCD corrections are roughly within
ten percent and not depend on $\tan \beta$ very much. However,
one will see soon that the effects, though only in ten percent,
will make substantial changes for
the constraints on the parameter space of 2HDM.
In order to see the experimental uncertainties
of the measurements of the branching ratio $BR(b\to s \gamma)$,
the top mass $m_t$ and the strong coupling constant
$\alpha_s(M_Z^2)$ how to affect the conclusions, we plot
the branching ratio $BR(b\to s \gamma)$ (the central value also with
three standard deviation upper and lower
bounds achieved by CLEO recently) versus $\tan \beta$
so as to show the constraints on $\tan \beta$ and on the mass of charged
Higgs $M_\phi$ for Model I (Fig.2a) and Model II
(Fig.2b), and plot the bends in the parameter space $\tan \beta$
versus $M_\phi$ to show the constraints caused by taking
$m_t=176 GeV$ with $\alpha_s(M_Z^2)=0.12$ and $0.13$ (the dashed lines)
and $\alpha_s(M_Z^2)=0.125$ with $m_t=163 GeV$ and $189 GeV$ (the
solid lines) for Model I (Fig.3a) and Model II (Fig.3b)
respectively. Here it is interesting to note that
in Fig.3b, the dashed lines approach to $350 GeV$ and $390 GeV$ respectively
whereas the solid lines approach to $310 GeV$ and $430 GeV$ respectively
when $\tan \beta$ approaches to infinity.
All lines in Fig.3, are obtained by
taking the 95\% C.L. value of CLEO, $1.0\times 10^{-4}<Br(B\to X_s \gamma)
< 4.2\times 10^{-4}$\cite{cleo2}. It is shown that the parameter space is more
sensitive for changing of $m_t$ than for $\alpha_s$, especially in
Model II.

One may see from Figs.2a,3a that for Model I, there are two bands
in the $\tan \beta$-$M_\phi$ plane, excluded by our reanalysis with
the latest measurements on $b\to s+\gamma$ and $m_t$.
Finally, as another result of the reanalysis, in Fig.4 we plot the
$\tan \beta$ versus $M_{\phi}$ plane to show the excluded region
of the parameters for Model II at 95\% C.L.. The solid
line corresponds to the conclusion
of the present analysis with the central values of all the
parameters ($m_t, \alpha_s$ and the branching ratio of the
semileptonic decay of $b$ quark etc.);
the dashed line corresponds
to a possible parameter: $|V_{ts}^* V_{tb}|^2/|V_{cb}|^2=0.99$
instead of $|V_{ts}^* V_{tb}|^2/|V_{cb}|^2=0.95$\footnote{This value is
widely adopted for all the earlier analysis. In fact, there
are uncertainties from the measurements and the formulars on the branching
ratio of the inclusive semileptonic decay of $B$ meson $BR(B\to X_c
e\bar{\nu})$
(see eq.(11)) as well as those from the determination of the CKM matrix
elements. Concerning the facts of the uncertainties
and the possibility for more than three generations
of the elementary fermions, we try to change the value of
$|V_{ts}^* V_{tb}|^2/|V_{cb}|^2$, so as to let one see the tendency
for the constraints when the value becomes greater (equivalent to
that the value of $BR(B\to X_c e\bar{\nu})$ becomes smaller than $10.8\% $
etc. .}; and the dot-dashed line corresponds to the results
obtained by other authors\cite{s1}, which do not include the piece of
the QCD corrections from $m_t$ to $M_W$,

In conclusion, due to the QCD corrections from $m_t$ to $M_W$,
the new experimental value of $m_t$ and the bounds for
$b \to s \gamma$, the constraints for 2HDM are strained substantially.
For instance, the lower bound for the mass of the charged Higgs is put
up at least $150 GeV$ for Model II.

\vspace{4mm}
\noindent
{\Large\bf{ Acknowledgements}

\vspace{1ex}}

The work was supported in part by the National Natural Science Foundation
of China and the Grant LWTZ-1298 of Chinese Academy of Science.
The authors would like to thank Prof. T.D. Lee, the director of CCAST
for creating the very active institution in Beijing, since this work
actually begins in one of the domestic workshops (April-May, 1995)
of CCAST.

\newpage
\vspace{4mm}

\centerline{\large\bf{ Figure Captions}}

\vspace{1ex}

\begin{enumerate}

\item
Coefficients of operators $O_7$, $O_8$, as functions of $\tan\beta$
in Model I (a) and Model II (b). Solid and dashed lines correspond to
$C_7$ with and without QCD running from $m_t$ to $M_W$; dash-dotted and
dotted lines correspond to $C_8$ with and without QCD running from $m_t$
to $M_W$.

\item
Branching ratios of $b\to s\gamma$ depicted as functions of
$\tan \beta$, with different masses of the charged Higgs $M_\phi$
in Model I (a), and Model II (b).
The CLEO experiment's central value, upper and lower limit of 95\% C.L.
for this decay are also depicted as solid lines.

\item
$\tan \beta$-$M_{\phi}$ plane to show excluded region of parameters
of Model I (a), Model II (b). Every two solid lines denote the allowed
and excluded region(s) with $\alpha_s(m_Z)=0.125$  and $m_{top}=189 GeV$
or $m_{top}=163 GeV$; Every two dashed lines correspond to that of
$m_=176 GeV$ and $\alpha_s(m_Z)=0.13$  or $\alpha_s(m_Z) =0.12$.

\item
$\tan \beta$-$M_{\phi}$ plane to show excluded region of parameters
of Model II. Solid lines corresponds to all central value of parameters;
dashed line correspond to $|V_{ts}^* V_{tb}|^2/|V_{cb}|^2=0.99$; and dot-dashed
line correspond to results without QCD Corrections from $m_t$ to $M_W$,
obtained by other authors.

\end{enumerate}

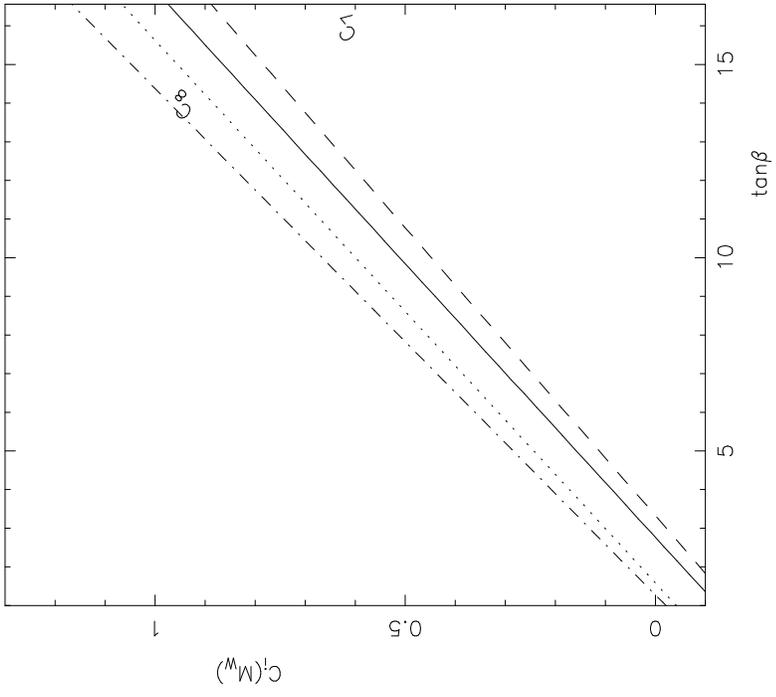
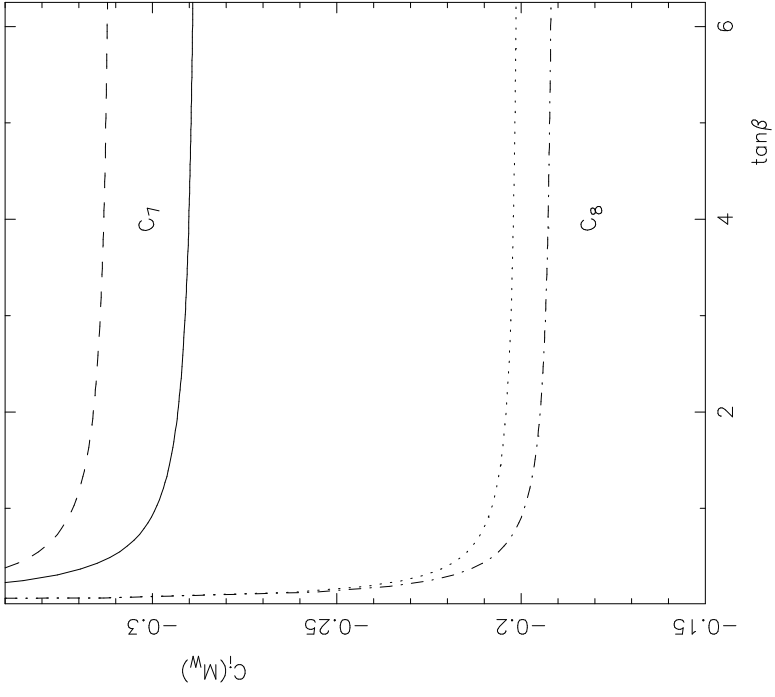
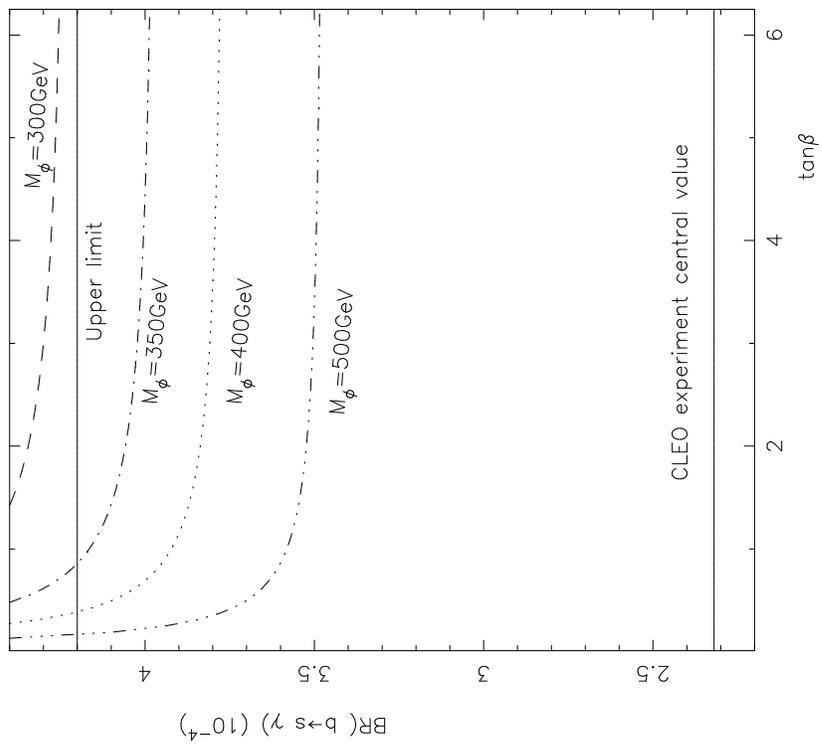
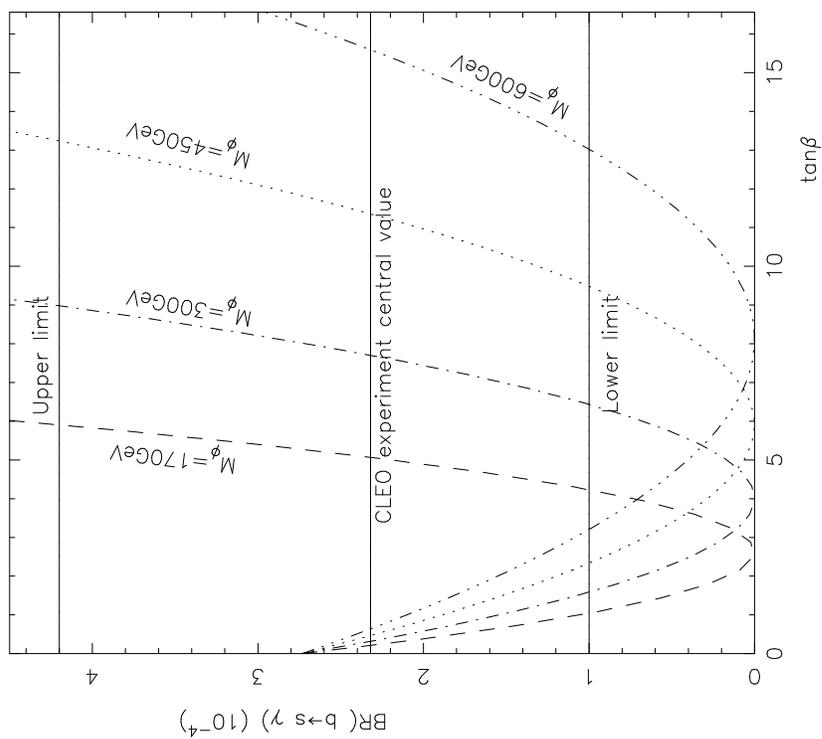
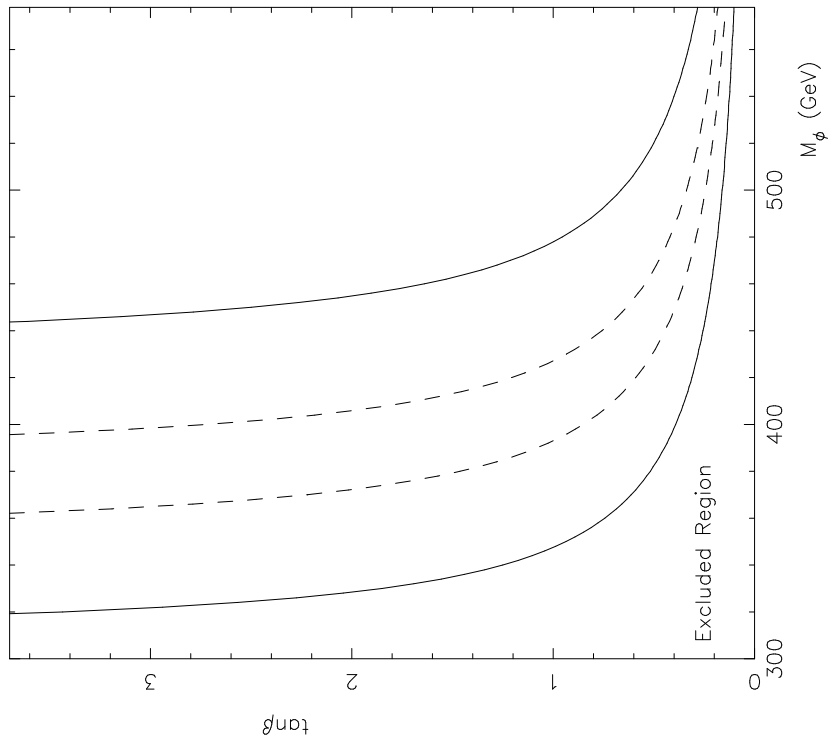
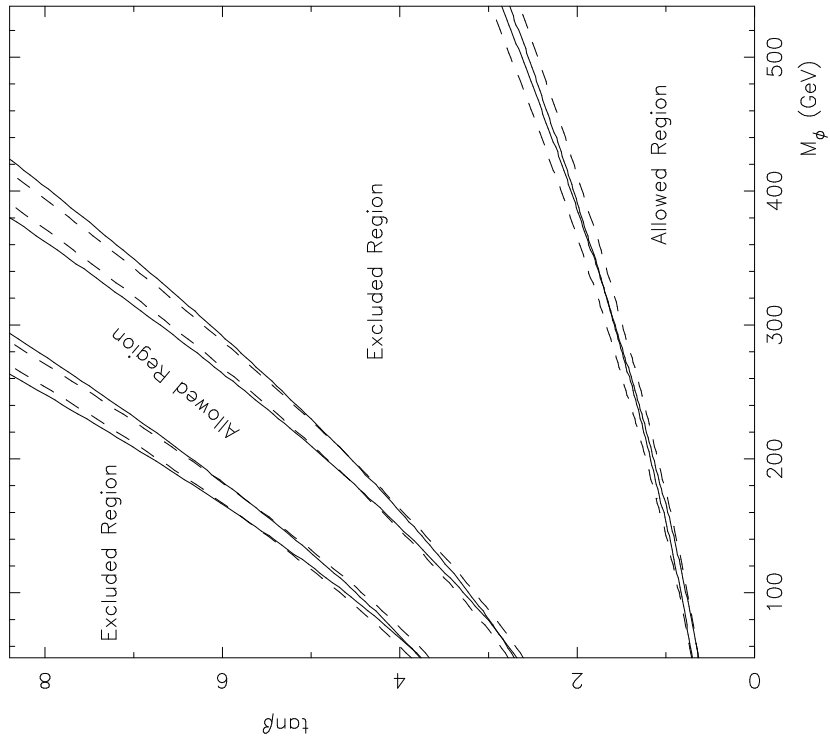
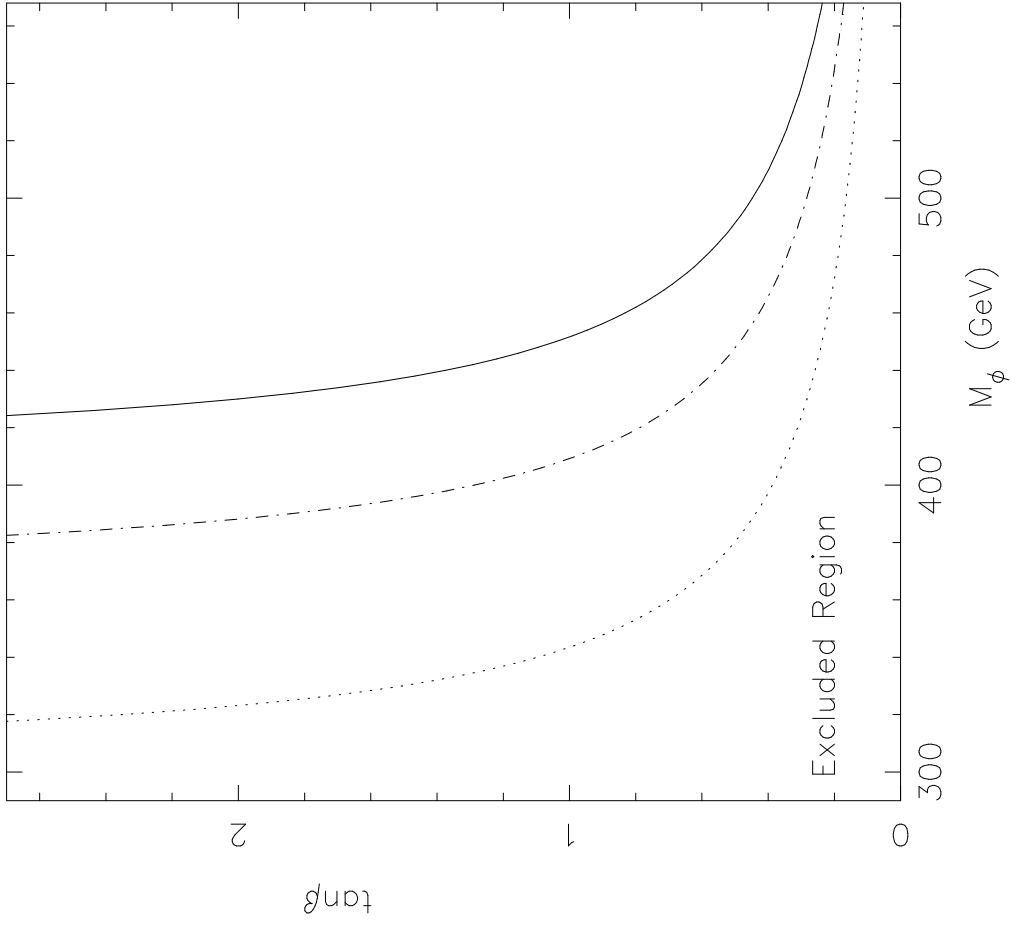
\end{document}